**Exploring the effects of numerical methods and slope limiters in heliospheric modeling**

*Tinatin Baratashvili[1](tinatin.baratashvili@kuleuven.be), Christine Verbeke[1,2], Rony Keppens[1], Stefaan Poedts[1,3]*

[1]*Centre for mathematical Plasma Astrophysics, KU Leuven, Leuven, Belgium*

[2]*Royal Observatory of Belgium, Ukkel, Belgium*

[3]*Institute of Physics, University of Maria Curie-Skłodowska, Lublin, Poland*

**Abstract**

Coronal mass ejections (CMEs) are large eruptions close to the solar surface, where plasma is ejected outwards into space at large speeds. When directed towards Earth, they interfere with Earth's magnetic fields and cause strong geo-effective storms. In order to mitigate the potential damage, forecasting tools are implemented. Recently, a novel heliospheric modelling tool, Icarus, has been implemented, which exploits the open-source framework MPI-AMRVAC as its core MHD solver. This new model efficiently performs 3D MHD simulations of the solar wind and the evolution of interplanetary CMEs with the help of advanced techniques, such as adaptive mesh refinement and gradual radial grid stretching. The numerical methods applied in the simulations can have significant effects on the simulation results and on the efficiency of the model. In this study, the effect of different combinations of numerical schemes and slope limiters, for reconstructing edge-based variables used in fluxes, is considered. We explore frequently exploited combinations from the available numerical schemes in MPI-AMRVAC: TVDLF, HLL and HLLC along with the slope limiters `woodward', 'minmod', 'vanleer', and 'koren'. For analysis purposes, we selected one particular solar wind configuration and studied the influence on variables at 1 AU in the equatorial plane. The goal is to find the optimal combination to produce accurate results fast and in a robust way so that the model can be reliable for day-to-day use by space weather scientists. As a conclusion, the best result assessed with these two criteria is the combination of the TVDLF scheme with the 'woodward' limiter.

**Introduction**

Space weather is a prevailing branch of physics that studies the time varying conditions near the Earth and in the inner heliosphere. It is affected by the solar activity, energetic particles resulting from flares and coronal mass ejections (CMEs). CMEs are massive magnetized plasma clouds (up to $10^{13}$ kg) that are ejected outwards from close to the solar surface into the lower solar corona and propagate in the heliosphere, disturbing Earth environment along the way (Gopalswamy et al., 2017). While propagating in the heliosphere, they interact with the ambient solar wind, which causes deformation, deflection and erosion. Their speeds can range from 100,000 m s$^{-1}$ to 3,000,000 m s$^{-1}$, with the average speed of ~ 450,000 m s$^{-1}$ (Webb et al., 2006). When they are directed towards Earth, they can cause geomagnetic storms when interacting with the Earth's magnetic field. Recently, in February 2022, 38 out of 49 SpaceX Starlink satellites suffered from such a minor geomagnetic storm. When the CMEs have strong magnetic field, their impact can even hinder the navigation or telecommunication systems, disrupt power systems, etc. On 13 March 1989, for instance, a strong CME hit Earth and soon after the Hydro-Quebec power grid failed, causing 9 hours of total blackout. The socio-economic loss due to such space weather events is large and as society depends on technology ever more, the extent of the possible damage also increases with time. In order to mitigate the consequences, physics-based forecasting tools are implemented. Sun-to-Earth modelling is challenging because the various physical phenomena are complex and difficult to model. As a result, and also to save CPU time, space weather forecasting procedures often involve different models that are coupled together, like coronal and heliospheric models. The coupling usually takes place in the 20-30 $R_\odot$ range (Narechania et al., 2021), beyond the radial distance where the wind becomes supersonic so that boundary conditions are simpler to enforce. One such operational tool is the EUropean Heliospheric FORecasting Information Asset (EUHFORIA; Pomoell & Poedts, 2018). EUHFORIA involves a combination of a semi-empirical (Wang-Sheely-Arge-like, Arge et al., 2000) coronal model and a physics-based 3D magnetohydrodynamics (MHD) heliospheric model. A similar popular operational tools are ENLIL(Odstrcil et al., 2004) and Space-weatherforecast-Usable System Anchored by Numerical Operations and Observations – Coronal Mass Ejection model (SUSANOO-CME; Shiota & Kataoka 2016). EUHFORIA, ENLIL and SUSANOO-CME, all apply the ideal 3D MHD equations to model the solar wind and then inject the CMEs from the inner heliospheric boundary at 0.1 AU (21.5 solar radii). Alternative models also simulate the immediate solar surroundings with physics-based models. The Alfvén-wave driven Solar Wind Model (AwSoM; Sokolov et al., 2013), for instance, starts at the chromosphere and includes the transition region. Another global corona model is the Magnetohydrodynamic Algorithm outside a Sphere (MAS; Linker et al., 1999) which solves the global 3D MHD equations including source and loss terms to model the heating and the losses due to radiation and thermal conduction. Recently, a new heliospheric wind and CME evolution model was implemented within the framework of MPI-AMRVAC (Xia et al., 2018, Keppens et al., 2021). This new solar wind and CME propagation tool is called Icarus (Verbeke et al., 2022) which solves the partial differential equations of ideal MHD in a frame that is co-rotating with the Sun, in order to obtain a steady background solar wind after an MHD relaxation phase. The aim of this new tool is to perform accurate and optimized simulations of CME evolution. This is achieved by advanced techniques, also explained in Baratashvili et al., (2022). The efficiency in terms of wall-clock time needed for the simulations is extremely important from the forecasting point of view, together with the accuracy of the results. On the other hand, shocks need to be captured as accurately as possible, avoiding numerical dissipation, to be able to use the shock information in models for accurate CME arrival time and for particle acceleration and transport, like PARADISE (Wijsen, 2020). In order to model the shocks associated with CMEs or co-rotating interaction regions (CIRs) in the domain, and the arrival time and the strength of CME shocks at Earth, it is crucial to choose the optimal numerical methods. Different numerical schemes are suited for different applications, and in this case we need to take into account the complexity of the magnetized solar wind interacting with the propagating CMEs. In order to obtain the most optimal numerical setting, different numerical methods and slope limiters were combined.

**Materials and methods**

The heliospheric simulations are performed with Icarus. The numerical domain of the heliospheric model is a spherical shell extending from 0.1AU to 2AU, including the orbit of Mars. It spans the full 360º in the longitude and 120º in the latitudinal direction (± 60º from the equatorial plane), avoiding the singularities at the poles (a spherical coordinate system is used). We consider different grid resolutions and name them the low, middle and high resolution. The characteristic cell sizes for each resolution can be found in Table 1.

|  | Radial [$R_\odot$] | Angular [degree] |
| --- | --- | --- |
| Low | 1.37 | 3.75 |
| Middle | 0.685 | 1.875 |
| High | 0.3425 | 0.9375 |

*Table 1. Cell sizes for Low, Middle and High resolution in Icarus. The radial cell size is given in solar radii, the longitudinal cell sizes are given in degrees and are the same as the latitudinal cell sizes.*

Advanced techniques such as AMR and grid stretching are available in this model, inherited from the general purpose AMR facilities in MPI-AMRVAC (Keppens et al., 2012, Xia et al., 2018, Keppens et al., 2021). Currently, only a simple, basic CME model is available in Icarus. This `cone CME model' represents a hydrodynamic (i.e. non-magnetized) plasma cloud with a homogeneous interior that is injected into the magnetized wind flow. The details of the CME injection in Icarus are given in Verbeke et al., (2022). For this study, we consider a particular solar wind configuration generated by the GONG (Global Oscillation Network Group) magnetogram corresponding to time 2012-07-12T11:54:00 and focus on modelling the background solar wind alone. The plasma variables at 0.1AU, obtained from the WSA corona model, are used as the inner boundary values for the heliosphere in Icarus and are radially extended to 2AU. This initial MHD state is then relaxed for 14 days, which takes only a few minutes of simulation time, after which a steady state is obtained.

Icarus is implemented within the framework of MPI-AMRVAC (Xia et al., 2018), a heavily parallelized code solving partial differential equations written as (near-) conservation laws. It is well-suited for magnetohydrodynamics applications, as the ideal MHD equations can indeed be formulated as conservation laws. Since numerous different problems have already been addressed with this code, many different numerical schemes are available. The documentation of MPI-AMRVAC (see http://amrvac.org) discusses the different spatial discretization methods and their suitable applications. In heliospheric simulations, the CIR and CME shocks need to be captured. The shocks are generated by the interactions of the high and slow speed streams, and by the fast CME propagation in the domain. In order to test how different numerical methods can model the shocks in the given application, three different numerical methods were chosen. The first method is a Total Variation Diminishing Lax-Friedrichs (TVDLF; G. Tóth and D. Odstrčil, 1996) scheme. This TVDLF (also called Local Lax Friedrichs) scheme is a very robust numerical scheme, easily used for any set of hyperbolic partial differential equations, and its TVD character makes it monotonicity preserving. The TVDLF scheme is a second-order scheme, both in time and space, that does not use a Riemann type solver. The scheme is rather robust and fast, but more diffusive than other second-order schemes. Other schemes we are considering are the HLL (Harten et al., 1983) and HLLC (Toro, Spruce & Speares, 1994) methods. HLL and HLLC are approximate Riemann type solvers, which make further approximations in their corresponding representation of the Riemann fan. HLL uses only two wave speeds, while HLLC includes the contact discontinuity as well. Usually, the HLL representation behaves similarly to the TVDLF scheme, with minor improvements. The three mentioned second-order numerical schemes are used in combination with different slope limiters in the cell center to cell-face reconstructions, and they will numerically affect the steep gradients caused by shocks in the domain. The default, robust limiter in the MPI-AMRVAC documentation is called 'minmod' (Yee et al., 1989). It is also addressed as 'MINBEE' or 'MINA' limiter. This is a classic second-order symmetric TVD limiter and considered as one of the most diffuse limiters. Another limiter we are considering, is a 'woodward' limiter (van Leer, 1977), which is also a second-order limiter. In the literature, it is also referred to as 'monotonized central (mc)' limiter. The next considered limiter for the present study is similar to the 'woodward' limiter, namely the 'vanleer' limiter (van Leer, 1974). Finally, the last limiter we are considering is the third-order asymmetric limiter 'koren' (Koren, 1993). This limiter is slightly more diffuse than other third-order limiters. The goal of this study is to compare all the combinations of the described numerical methods and limiters with different grid resolutions, in order to obtain the most detailed and most efficient numerical model.

**Results**

We considered all 36 different combinations of the 3 different grid resolutions (low, middle high), the 3 mentioned numerical methods (HLL, HLLC and TVDLF) and the 4 slope limiters ('minmod', 'vanleer', 'woodward', 'koren'). Table 2 shows the simulation wall-clock times for the twelve middle resolution simulations, as this is the standard and most often used resolution set-up in an operational setting. All the simulations are performed on 1 node with 36 processors on the Genius cluster at the Vlaams Supercomputing Center. Here, the simulations take the relaxation period of 14 days into consideration and also a forecast time window of 10 days, i.e. 24 simulated days in total. The low and high resolution simulation timings are given in Appendix A.

| Middle resolution | 'minmod' | 'woodward' | 'vanleer' | 'koren' |
|---|---|---|---|---|
| TVDLF | 1h 47m | 1h 47m | 1h 47m | 2h 6m |
| HLL | 1h 47m | 2h 6m | 2h 6m | 2h 12m |
| HLLC | 2h 1m | 2h 17m | 2h 17m | 2h 40m |

*Table 2. The simulation wall-clock times for the twelve combinations of the three numerical methods and the four slope limiters performed for the middle resolution simulations.*

From Table 2 we can see that among the three considered numerical methods, TVDLF produces the fastest results. Among the slope limiters, 'minmod' yields the fastest simulations, while the 'vanleer' and 'woodward' limiters behave similarly and the 'koren' limiter is the slowest. The HLLC numerical scheme yields the longest times to perform the simulations.

In order to assess the accuracy of each scheme-limiter combination, first we consider the behavior of the different limiters for a fixed numerical method. Below, we demonstrate the results in combination with the TVDLF scheme; the results for the HLL and HLLC schemes are given in Appendix B.

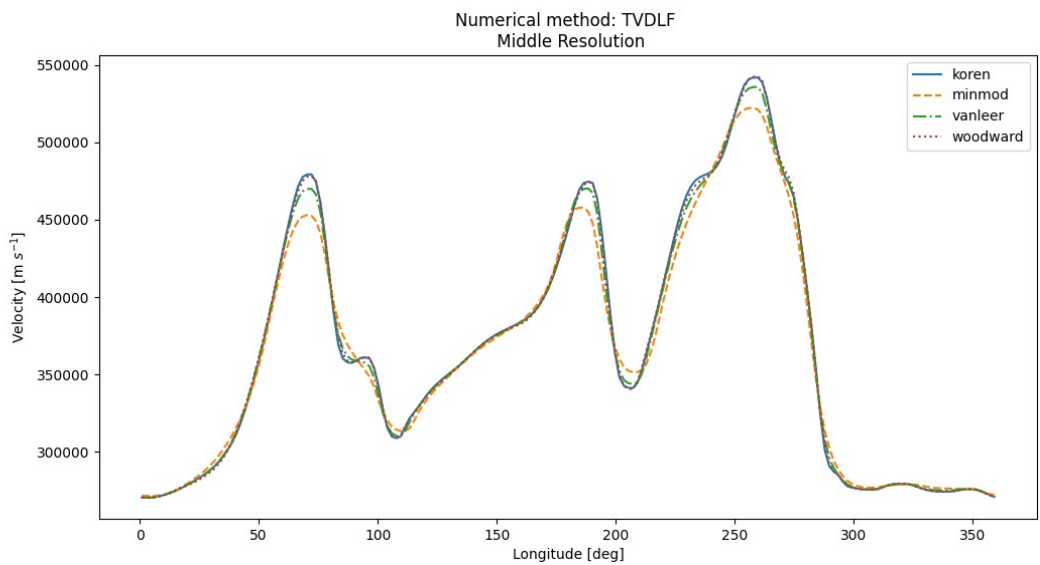

Figure 1. *Velocity values at 1 AU in the equatorial plane. The horizontal axis shows the longitudes in degrees, while the vertical axis shows the velocity in [m s$^{-1}$]. The results are plotted for the 4 different slope limiters in combination with the TVDLF scheme on the middle resolution grid.*

From Figure 1 we can see that 'minmod' indeed produces the smoothest results, as expected, followed by the simulation using the 'vanleer' limiter. The sharpest profiles are obtained with the 'woodward' and 'koren' limiters. The same behavior is observed for the simulations in combinations with HLL and HLLC shown in appendix B.

Next, we compare the three different numerical schemes in combination with the 'woodward' limiter. The combinations with the 'minmod', 'vanleer' and 'koren' limiters are given in Appendix C.

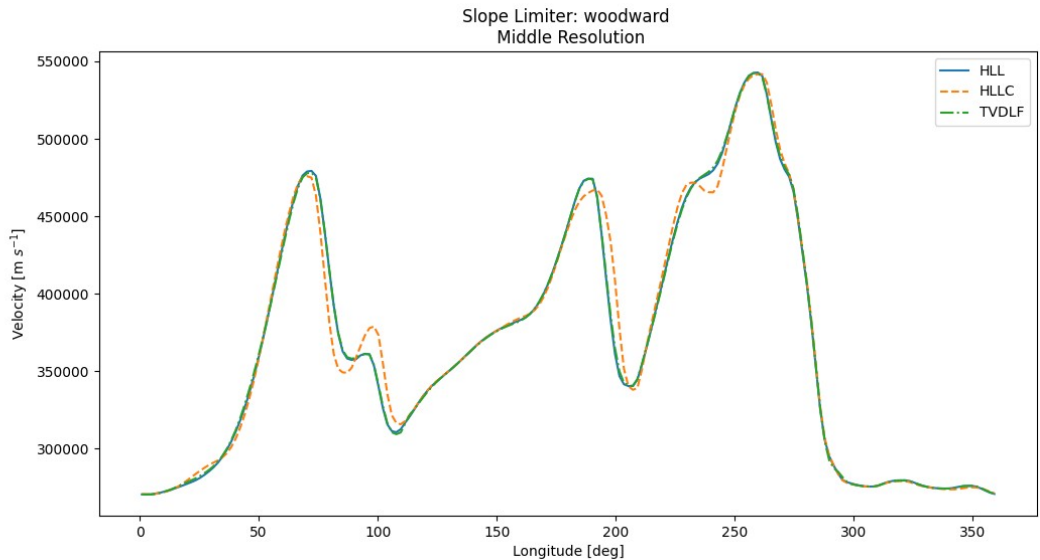

Figure 2. *Velocity values at 1 AU in the equatorial plane. The horizontal axis shows the longitudes in degrees, while the vertical axis shows the velocity in [m s$^{-1}$]. The results are plotted for the 3 different numerical methods in combination with the 'woodward' limiter on the middle resolution grid.*

Figure 2 compares the accuracy of the different numerical methods in combination with the 'woodward' slope limiter. We can see that HLL and TVDLF, given by the blue and green curves, respectively, produce very similar results. The results modelled by the HLLC scheme seems to be sharper, resolving more variation, especially at the areas where the speed of the wind is changing significantly. A similar behavior is spotted from the simulations in combinations with the other limiters shown in appendix C.

Heliospheric modelling is usually interesting to study the magnetic field in the Earth's surroundings, or the CME properties upon the arrival at Earth. In the following simulations, we decided to fix the numerical scheme to TVDLF and check how different limiters affect first the magnetic field (Figure 3) and then the CME features upon arrival at 1AU (Figure 4).

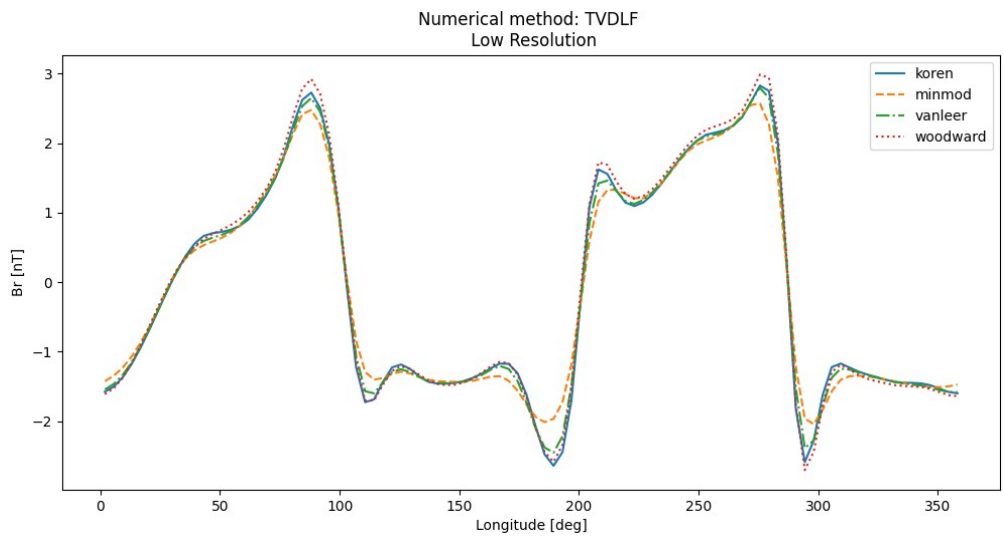

Figure 3. *Radial component of the magnetic field values at 1 AU in the equatorial plane. The horizontal axis shows the longitudes in degrees, while the vertical axis shows the $B_r$ in [nT]. The results are plotted for the 4 different slope limiters in combination with the TVDLF scheme on the low resolution grid.*

Figure 3 shows how the different limiters model the magnetic field. It is notable that the behavior is similar to what was observed when comparing the radial velocity data. Again, 'woodward' and 'koren' produce the sharpest profiles, followed by the 'vanleer' limiter and the smoothest profiles are given in the simulations with the 'minmod' limiter.

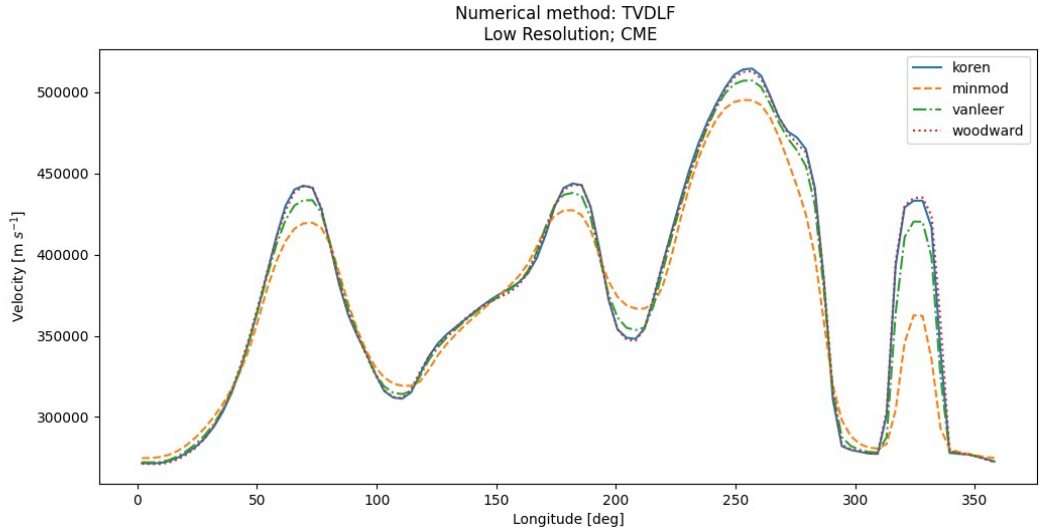

Figure 4. *Velocity values at 1 AU in the equatorial plane. The horizontal axis shows the longitudes in degrees, while the vertical axis shows the velocity in [m s$^{-1}$]. The CME arrives at 1AU between longitudes 300º and 350º. The results are plotted for the 4 different slope limiters in combination with the TVDLF scheme on the low resolution grid.*

Figure 4 shows the velocity profiles at 1AU for the different limiters. In this, simulations the CME is also modelled with a simple cone CME model similar to Verbeke et al. (2022). The CME shock is best modelled by the 'woodward' and 'koren' limiters. The 'vanleer' limiter produces less sharp profiles and the 'minmod' limiter models the smallest shock.

**Discussion**

The results provided in the previous section provide a better insight in the efficiency of the different combinations of numerical methods and slope limiters. In order to choose the most optimal and 'default' combination for the heliospheric simulations in Icarus, the following criteria have been considered: i) how detailed is the modelled data and ii) the wall-clock time of the simulations.

Table 1 shows that the simulation is the fastest in the following combinations: TVDLF + 'minmod', 'woodward' or 'vanleer', and HLL + 'minmod'. In all these cases, the simulations on the middle resolution grid are finished in under 2 hours. From Figure 1 we compare the performance of different limiters. From this comparison, it is clear that the 'woodward' and 'koren' limiters are the most favourable ones, as they show more detailed profiles with better resolved gradients. And finally, Figure 2 compares the different numerical methods. From this figure, it can clearly be seen that the HLLC scheme shows the best resolved results. If we put aside the operational factor, i.e. the CPU time consumption, the best results are given by the combinations of the HLLC numerical method in combination with the 'woodward' or 'koren' slope limiters. However, as we need to take into consideration also how efficient the simulations can perform, the more optimal choice would be the combination of the TVDLF numerical method with the 'woodward' slope limiter. Following the main study of this paper, we performed the comparison for 2 most interesting profiles: the magnetic field modelled at 1AU (Figure 3) and the CME arrival au 1AU (Figure 4). From these figures, the combination of the TVDLF numerical method and the 'woodward' limiter produced the sharpest profiles. This combination is thus chosen as the default setting in Icarus in order to obtain the most optimized simulation setting. However, because the MPI-AMRVAC framework provides the freedom to select the numerical schemes and limiters with the minimal implementation costs, depending on the purpose of the simulation, the combinations can be changed, taking into consideration the numerical details (number of ghost cells, order of stepping in time, etc.). For the purpose of studying small-scale structures in the heliosphere or the SEP modelling, the high resolution domain can be combined with the HLLC numerical method. The simulation setup can be easily modified in Icarus, but as a result of this study, combining different resolution grids, numerical methods and slope limiters, the most optimal combination for solar wind simulations is achieved with the TVDLF scheme with the 'woodward' limiter.

**Acknowledgements**

This research has received funding from the European Union's Horizon 2020 research and innovation programme under grant agreement No 870405 (EUHFORIA 2.0) and the ESA project "Heliospheric modeling techniques" (Contract No. 4000133080/20/NL/CRS). These results were also obtained in the framework of the projects C14/19/089 (C1 project Internal Funds KU Leuven), G.0D07.19N (FWO-Vlaanderen), SIDC Data Exploitation (ESA Prodex-12), and Belspo project B2/191/P1/SWiM. The Computational resources and services used in this work were provided by the VSC-Flemish Supercomputer Center, funded by the Research Foundation Flanders (FWO) and the Flemish Government-Department EWI. RK acknowledges funding from the European Research Council (ERC) under the European Union's Horizon 2020 research and innovation program (grant agreement No. 833251 PROMINENT ERC-ADG 2018).

**Appendix A**

| Low resolution | 'minmod' | 'woodward' | 'vanleer' | 'koren' |
| --- | --- | --- | --- | --- |
| TVDLF | 24m | 24m | 24m | 24m |
| HLL | 29m | 29m | 29m | 29m |
| HLLC | 36m | 36m | 36m | 36m |

*Table A.1. The simulation times for the combinations of the numerical methods and slope limiters performed on the low resolution computational domain.*

| High resolution | 'minmod' | 'woodward' | 'vanleer' | 'koren' |
|---|---|---|---|---|
| TVDLF | 12h 45m | 13h 24m | 13h 56m | 16h 1m |
| HLL | 15h 43m | 16h 4m | 15h 18m | 17h 47m |
| HLLC | >20h | 19h 19m | 18h 59m | >20h |

*Table A.2. The simulation times for the combinations of the numerical methods and slope limiters performed on the high resolution computational domain.*

**Appendix B**

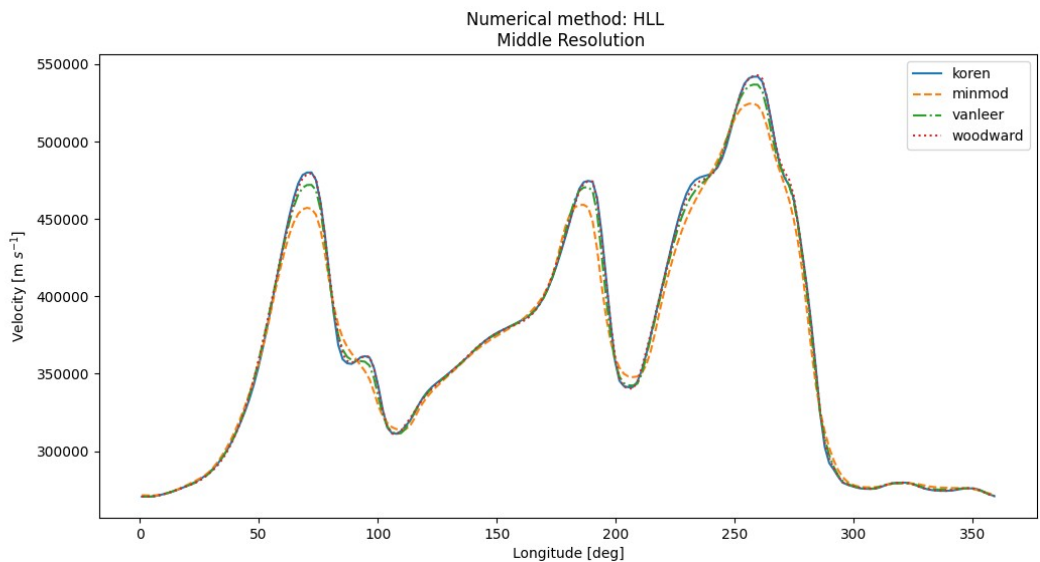

*Figure B.1. Velocity values at 1 AU in the equatorial plane. The horizontal axis shows the longitudes in degrees, while the vertical axis shows the velocity in [m s$^{-1}$]. The results are plotted for the 4 different slope limiters in combination with the HLL scheme on the middle resolution grid.*

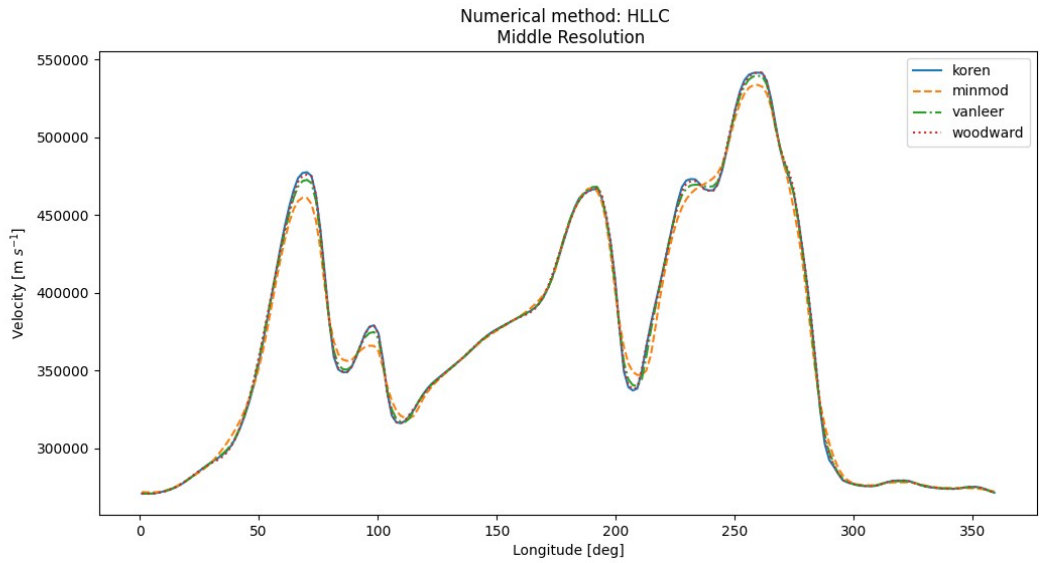

***Figure B.2.*** *Velocity values at 1 AU in the equatorial plane. The horizontal axis shows the longitudes in degrees, while the vertical axis shows the velocity in [m s$^{-1}$]. The results are plotted for the 4 different slope limiters in combination with the HLLC scheme on the middle resolution grid.*

**Appendix C**

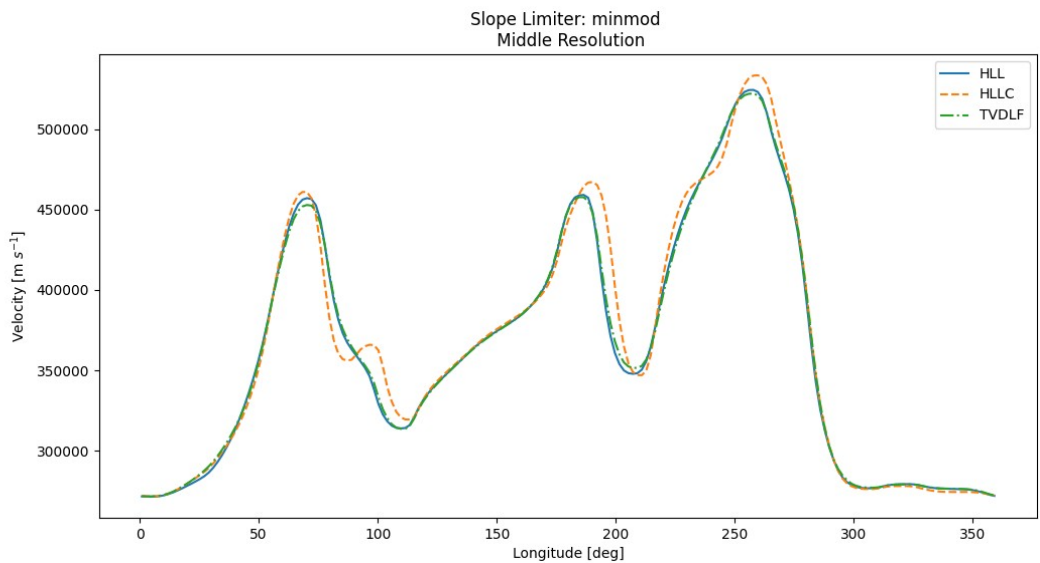

***Figure C.1***. *Velocity values at 1 AU in the equatorial plane. The horizontal axis shows the longitudes in degrees, while the vertical axis shows the velocity in [m s$^{-1}$]. The results are*

*plotted for the 3 different numerical methods in combination with the 'minmod' limiter on the middle resolution grid.*

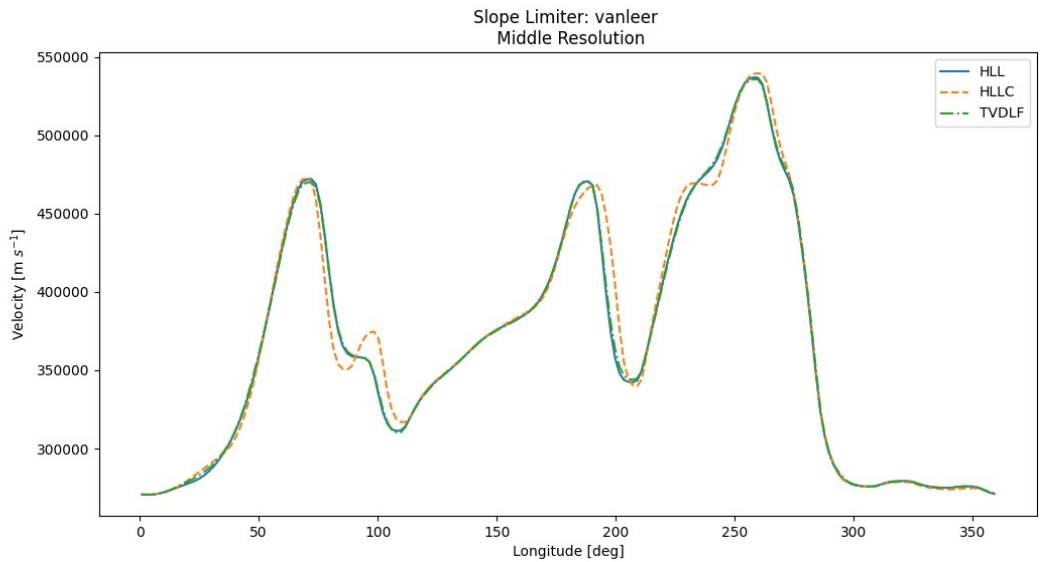

***Figure C.2**. Velocity values at 1 AU in the equatorial plane. The horizontal axis shows the longitudes in degrees, while the vertical axis shows the velocity in [m s$^{-1}$]. The results are plotted for the 3 different numerical methods in combination with the 'vanleer' limiter on the middle resolution grid.*

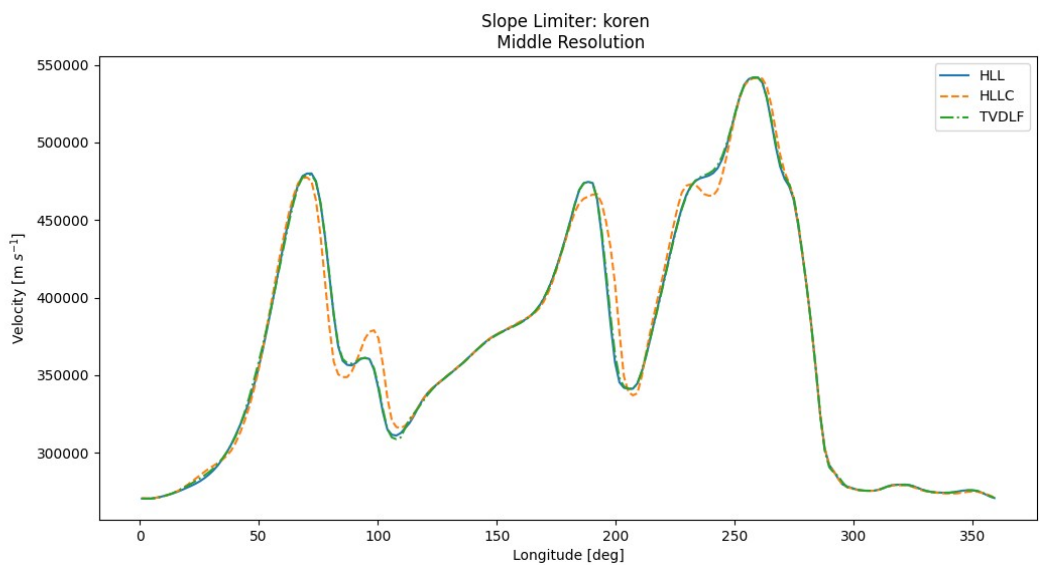

***Figure C.3.** Velocity values at 1 AU in the equatorial plane. The horizontal axis shows the longitudes in degrees, while the vertical axis shows the velocity in [m s$^{-1}$]. The results are*

*plotted for the 3 different numerical methods in combination with the 'koren' limiter on the middle resolution grid.*